\documentclass[12pt]{article}
 
\usepackage{moreverb,xspace,bm}
\usepackage{amsmath}
\usepackage{amssymb}
\usepackage{amsthm}
\usepackage{natbib}
\usepackage[colorlinks,bookmarksopen,bookmarksnumbered,citecolor=red,urlcolor=red,breaklinks=true]{hyperref}

 \usepackage{tikz}
 \usepackage{rotating}
 \usepackage{float,subfig}
 \usepackage{multicol}
\usepackage{graphicx}
 \usetikzlibrary{arrows,shapes,calc}
 \usetikzlibrary[positioning]
 \usetikzlibrary{patterns}

\newcommand{\ie}{\textit{i.e.\/}\xspace}
\newcommand{\eg}{\textit{e.g.\/}\xspace}

\newcommand{\bei}{\begin{itemize}}
\newcommand{\eei}{\end{itemize}}
\newcommand{\be}{\begin{equation}}
\newcommand{\ee}{\end{equation}}
\newcommand{\ben}{\begin{enumerate}}
\newcommand{\een}{\end{enumerate}}
\newcommand{\bb}[1]{\begin{block}{#1}} 
\newcommand{\eb}{\end{block}} 

\usepackage{enumerate}
\usepackage{natbib}
\usepackage{url} 

\newcommand{\blind}{1}

\newcommand{\cip}{\mbox{$\perp\!\!\!\perp$}}
\newcommand{\nip}{\not\!\!\!\,\cip}

\begin{document}

\def\spacingset#1{\renewcommand{\baselinestretch}%
{#1}\small\normalsize} \spacingset{1}


\if1\blind
{
  \title{\bf  Bayesian modelling for binary outcomes in the Regression Discontinuity Design}
 \author{Sara Geneletti \\
    Department of Statistics, London School of Economics and Political Science \vspace{15pt}\\    
    Federico Ricciardi, Aidan O'Keeffe and Gianluca Baio\\
    Department of Statistical Science, University College London
    }
  \maketitle
} \fi

\if0\blind
{
  \bigskip
  \bigskip
  \bigskip
  \begin{center}
    {\LARGE\bf Bayesian modelling for binary outcomes in the Regression Discontinuity Design.}
\end{center}
  \medskip
} \fi

\bigskip
\begin{abstract}
The Regression Discontinuity (RD) design is a quasi-experimental
design which emulates a randomised study by
exploiting situations where treatment is assigned according to a
continuous variable as is common in many drug treatment guidelines. 

The RD design literature focuses principally on continuous
outcomes. In this paper we exploit the link between the RD design and
instrumental variables to obtain a causal effect estimator, the risk
ratio for the treated (RRT), for the RD design when the outcome is
binary.

Occasionally the RRT estimator can give negative lower confindence bounds. In the Bayesian framework we impose prior constraints that prevent this from happening. This is novel and cannot be easily reproduced in a frequentist framework.

We compare our estimators to those based on estimating equation and generalized methods of moments methods. Based on extensive simulations our
methods compare favourably with both methods.

We apply our method on a real example to estimate the effect of statins on the probability
of Low-density Lipoprotein (LDL) cholesterol levels reaching recommended levels.

\end{abstract}

\noindent%
\vfill

\newpage
\spacingset{1.45} 

\section{Introduction}

The Regression Discontinuity (RD) design is a quasi-experimental
design introduced in the 1960's in \citet{ThistlethwaiteCampbell1960}
and widely used in economics and related social sciences
(\cite{ImbensLemieux2008}) and more recently in the
medical sciences (\cite{Genelettietal:2015,Boretal:2014}). The RD
design enables use of routinely gathered medical data from general
practice (family doctors) to evaluate the causal effects of drugs
prescribed according to well-defined decision rules. Most of the RD design literature focuses on continuous outcomes. Here we develop Bayesian approach for binary outcomes which are frequently of primary interest in health care contexts.

The RD design naturally leads to an Instrumental Variable (IV)
analysis and so we adapt the IV based Multiplicative Structural Mean
Model (MSMM) Risk Ratio for the Treated (RTT) estimator
(\cite{ClarkeWind:2012,ClarkeWindmeijer:2010,HernanRobins2006}) to
this context. The RRT is a measure of the change in risk for those who
received the treatment. The MSMM estimator for the RRT is known to
occasionally misbehave in that lower 95\% confidence intervals can be
negative (\cite{ClarkeWindmeijer:2010}). It is possible that this is
why this estimator is not widely used despite its relatively simple
formulation and partly what motivated the widespread use of
generalised methods of moment based estimator
(\cite{Clarkeetal:2015}). Naive Bayesian estimators of the MSMM RRT
also suffer from this problem; however we circumvent this issue by
imposing prior constraints that prevent the posterior MCMC sample from
dropping below zero. This is a novel implementation of prior
constraints and cannot be reproduced in a frequentist framework.

Once the problem of negative values is taken care of, the MSMM RRT
estimator turns out to be very flexible as we can estimate its
components using a large number of parametric and potentially
semi-parametric models. While we favour Poisson regression models due
to theoretical considerations (see Section \ref{sec:assumptions}) the
framework is not limited to this choice and we give examples of
alternatives in the Supplementary materials.

We also compare our estimators to those based on the generalized
methods of moments (\cite{Clarkeetal:2015}) amongst others. Based on
extensive simulations our methods compare favourably with the
frequentist estimators especially in circumstances that are not ideal.

The paper is organised as follows: In Section \ref{sec:rd} we briefly
describe the RD design and introduce our example. Section
\ref{sec:assumptions} lays out our notation and assumptions. In
Section \ref{sec:models} we describe in detail our models as well as
giving a short overview of the competing method based on the
generalized methods of moments. We present the results of a simulation
study in Section \ref{sec:sim}. Section \ref{sec:analysis} follows
with the results of the real application. We finish with some discussion in Section
\ref{sec:conclusion}.

\section{Background and example}
\label{sec:rd}

An analysis based on the RD design is appropriate for public health
interventions that are implemented according to pre-established
guidelines proposed by government institutions such as the Federal
Drug Administration (FDA) in the US and the National Institute for
Health and Care Excellence (NICE) in the UK. Specifically, when a
decision rule is based on whether a continuous variable exceeds a
certain threshold, it becomes possible to implement the RD design. In
our running example we use data from The Health Improvement
Network a UK data set. We investigate the effect of prescribing
statins, a class of cholesterol-lowering drugs, on reaching the NICE
recommended LDL cholesterol targets of below 2 or 3 mmol/L for healthy and high risk patients respectively. Between
2008 and 2014 NICE guidelines recommended that statins should be
prescribed to patients whose 10 year cardiovascular risk score
exceeded 20\%\, provided they had no history of cardiovascular
disease. If we are willing to assume that individuals just above and
below the 20\% threshold are exchangeable -- an essential condition
for causal inference in the RD design -- then we have a quasi-randomised
trial with those just below the threshold randomised to the
no-treatment ``arm'' and those just above randomised to the treatment
``arm''. Thus any jump or discontinuity in the values of the outcome
across the threshold can be interpreted as a local causal effect or
risk ratio in the case of binary outcomes. For more details see
\cite{Genelettietal:2015}.

There are two types of RD designs. The first is termed \textit{sharp}
and refers to the situation where the guidelines are adhered to
strictly. In our application we would encounter a sharp RD design if
all the doctors complied with the NICE guidelines and prescribed statins
exclusively to patients with a 10 year cardio-vascular risk score above
20\%. In practice this is often not the case, and our application makes no exception. There is some ``contamination'' whereby individuals who
are below the threshold are prescribed statins while other who are above the threshold receive no prescription. This situation is termed a \textit{fuzzy} RD design.


An open question in the RD design literature is how close to the
threshold units have to be in order for the RD design to be valid
(\cite{Imbens:2011} and \cite{CCT:2015}). 
The term bandwidth is
typically used to refer to the distance from the threshold within
which the units above and below are used for analysis. In line with
\cite{Genelettietal:2015} we estimate the RRT for four bandwidths of
varying size and assess sensitivity to these changes.

\subsection{Exploratory plots for the RD design}\label{sec:explot}

We first present plots (using subsets of the THIN data for the year 2008) that are typically used to justify the use of
the RD design in a fuzzy setting and are useful for understanding how the RD design works.  In all figures the red circles are individuals
who do not have statin prescriptions while blue x's represent those
who do have statin prescriptions. We plot the mean of the outcome
(continuous or binary) within bins of the risk score ($x$-axis)
against the risk score and fit a cubic spline. When the splines show
jumps at the threshold this indicates a discontinuity and thus a
potential causal effect at the threshold.

\begin{figure}[!h]
\begin{center}
\includegraphics[scale=0.5]{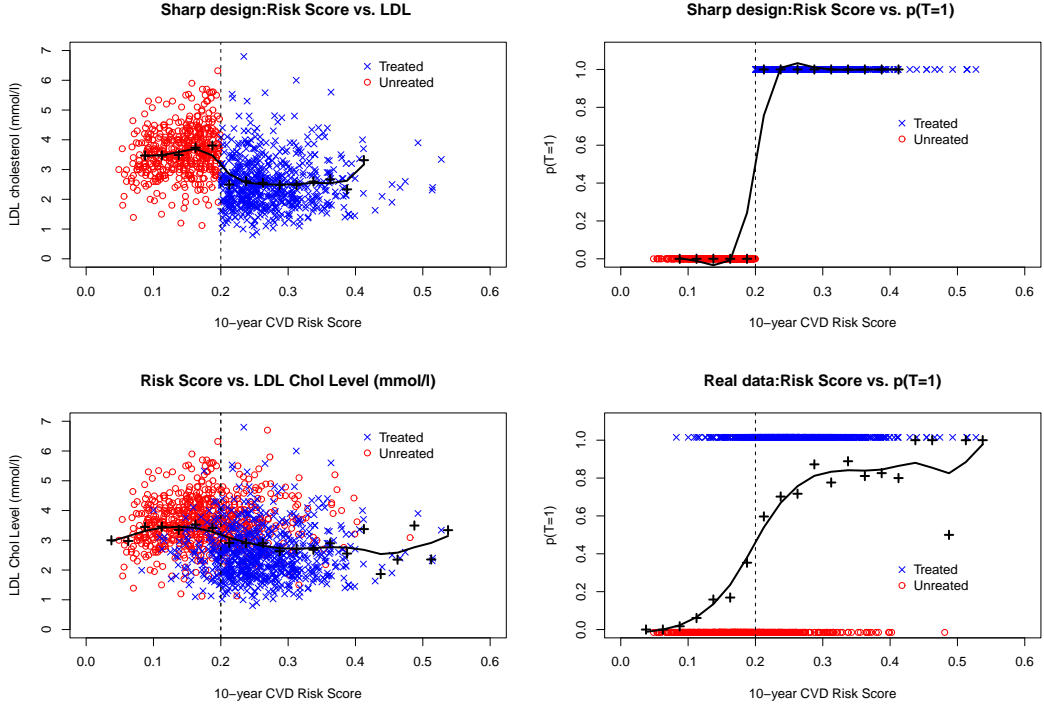}
\end{center}
\caption{\label{fig:ideal} Top row: The sharp design. Bottom row: The real (fuzzy) design. In both cases means within bins and corresponding fitted cubic spline are overlayed.}
\end{figure}

The top row of Figure \ref{fig:ideal} shows a sharp design. On the
left, the $y$-axis is a continuous measure of LDL cholesterol levels in
mmol/L. There is a small but noticeable downward jump at the
threshold. On the right is the corresponding graph of the raw
probability of treatment on the $y$-axis. Again there is a clear jump
from 0 to 1. On the bottom row is an example of a fuzzy design. This is clear as there are red
circles above the threshold and blue x's below. Despite the
fuzziness there is still a small downward jump at the threshold as
shown in the bottom left hand plot. The bottom right hand plot is the
corresponding raw probability plot. The increase of probability is
more gradual but there is a distinct jump at the threshold (\cite{Genelettietal:2015}).

\begin{figure}[!h]
\begin{center}
\includegraphics[scale=0.5,angle=-90]{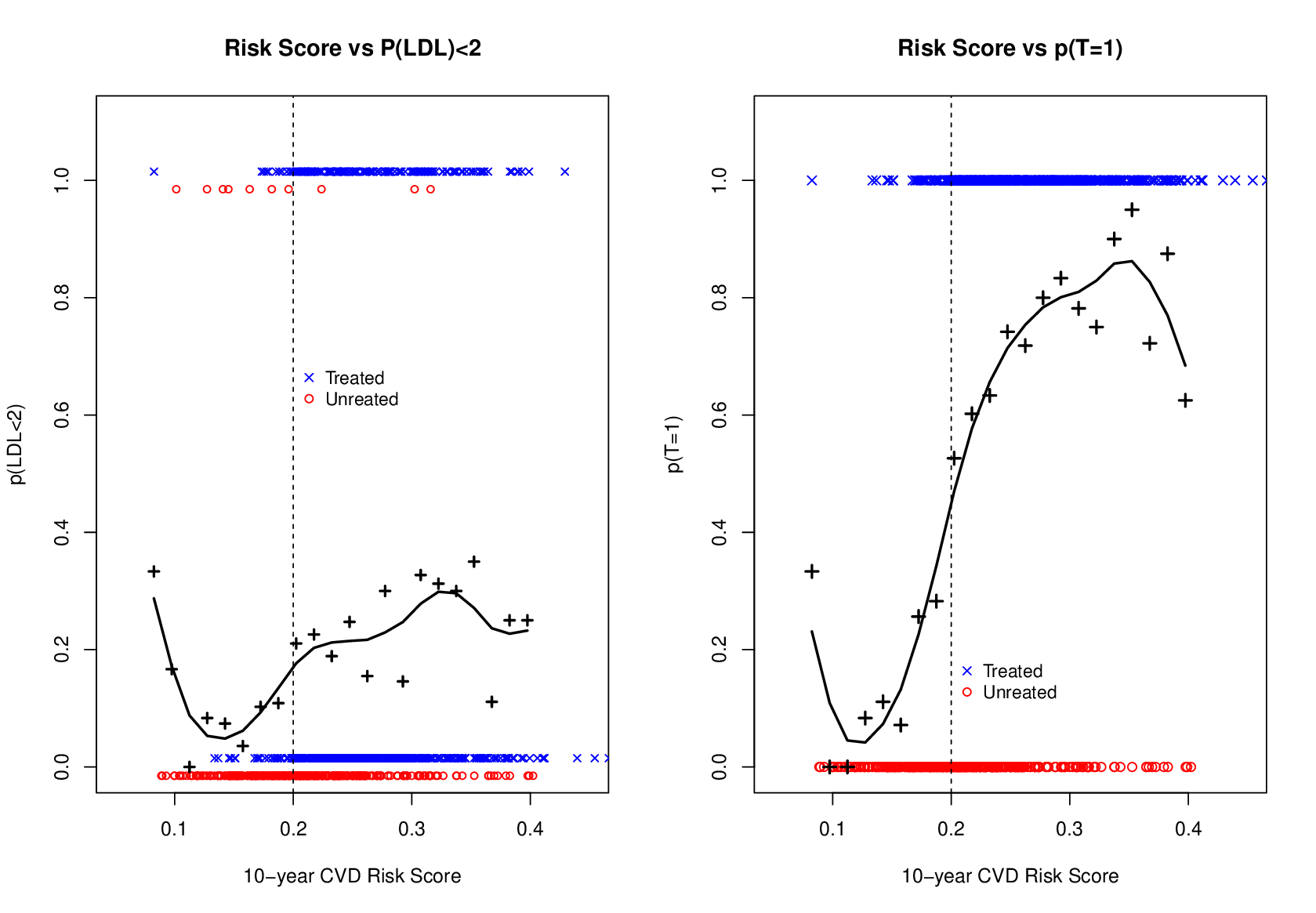}

\end{center}
\caption{\label{fig:binaryoutcome} Mean and probability plots for outcome LDL cholesterol going below 2 mmol/l. }
\end{figure}

Figure \ref{fig:binaryoutcome} shows the plots for the binary outcome
LDL cholesterol below 2 mmol/L which we analyse in Section
\ref{sec:analysis}. While there is no discernible jump in the outcome
(left), there is evidence in a jump in the probability of being
treated for such patients (right). Taken together, there is sufficient
support for a RD design analysis in this context. The corresponding
plots for LDL cholesterol dropping below 3 mmol/l are given in the
Supplementary Material and show a jumps in the outcome and the
probability of treatment.

\subsection{Assumptions and notation}
\label{sec:assumptions}

In order for the RD design to be appropriate a number of formal assumptions also
have to hold. These assumptions are expressed in different ways (\cite{HahnToddVDK2001,ImbensLemieux2008,vanderKlaauw2008,Lee2008}) and we give a brief overview of them as described in more detail in \cite{Genelettietal:2015,NayiaAidan:2015}. In the binary outcome case we need to make additional assumptions in order to identify the Risk Ratio for the treated our estimator of choice (\cite{Didelezetal:2010,HernanRobins2006}).


 We operate within a decision theoretic framework (\cite{Dawid:2002}),
 as this approach makes more explicit the assumptions needed to link
 causal (experimental) and observational quantities. However the RRT
 as identified in (\ref{eq:MSMMform}) below and consequently the
 Bayesian methods we apply do not rely on this framework and can be interpreted within a counterfactual or potential outcomes paradigm.

We express our assumptions in the language of conditional independence
following \cite{APD.79}. We say that if variable $A$ is independent of
another $B$ conditional on a third $C$, then $p(A,B\mid C)=p(A\mid
C)p(B\mid C)$ and we write $A \cip B\mid C$. We refer to our example
to anchor the theoretical arguments. Generalising to other contexts is
straightforward.

Let $X$ be the 10 year cardiovascular risk score. The threshold
indicator $Z$ is such that $Z=1$ if $X \geq 0.2$ and $Z=0$ if $X < 0.2$.
Let $T$ represent statins prescription (not whether the patient takes
the treatment); $T=1$ means statins are prescribed and $T=0$ means
they are not. Also, let $\bm{C}=\{\bm{O}\cup\bm{U}\}$ be the set of
confounders, where $\bm{O}$ and $\bm{U}$ indicate fully observed and
partially or fully unobserved variables, respectively. $Y$ is the
binary outcome variable where $Y=1$ if LDL cholesterol is below 2
(resp, 3) and 0 otherwise.

To reflect the fact that these assumptions are only valid around the
threshold, we assume throughout the paper an additional conditioning
on $X \in[0.2,0.2 + h]$ if above the threshold and $X \in
[0.2- h,0.2]$ below the threshold for some suitably small
$h$. We do not explicitly write this conditioning except where
necessary.

Finally, in order to be able to use data around the
threshold to estimate quantities with causal meaning we introduce a
regime indicator $\Sigma$ as in (\cite{NayiaAidan:2015}). $\Sigma$ represents which of the three possible regimes or situations we find ourselves in: the interventional regime (i.e. the RCT), the sharp RD design or the fuzzy RD design.
\[
    \Sigma= 
\begin{cases}
    t, & \mbox{for $T=t$ $(=0,1)$, under the interventional regime;}\\
    s, & \mbox{for the sharp RD design;}\\
    f, & \mbox{for the fuzzy RD design.}
\end{cases}
\]

When $\Sigma=t$ we mean that $T$ is set to $t$ with no uncertainty as
in an RCT with perfect compliance. In this case it is akin to the
``do'' operator (\cite{Causality}) or the intervention variable $F$ in
\cite{Dawid:2002}. For clarity we replace conditioning on
$\Sigma=\varsigma$ by a subscript on the expectation
$E_{\varsigma}(Y\mid X,C)$. It is clear from the definition of
$\Sigma$ that both RD designs are types of observational data. By
defining $\Sigma$ we can formalise, in terms of conditional
independences, when it is possible to make inference about causal
quantities from estimates based on observational data. We make two
further assumptions involving the regime $\Sigma$:
\begin{enumerate}
\item[A1.] $(C,X,Z)  \cip  \Sigma;$
\item[A2.] $Y  \cip  \Sigma  \mid (C,X,Z,T,   )$
\end{enumerate}

Assumption 
A1 says that the value of the
confounders, the assignment variable (and trivially the threshold
indicator) are marginally independent of the context in which they
arise, \ie\ in an experiment or within and RD design. Under 
A2, the value of the outcome does not depend on the
regime if we take into account the confounders, the assignment
variable (and trivially the threshold indicator) as well as the
prescription status. These assumptions are termed ``sufficiency''
assumptions (\cite{DawidDidelez:2010}) and allow us to move from the interventional to the
RD design regimes.

\subsection{RD design assumptions}\label{sec:rddassumptions}

The first three RD design assumptions are essentially IV assumptions whilst the fourth is specific to the RD design. See \cite{Genelettietal:2015} for more details and interpretation.
\begin{enumerate}
\item[R1.] \textit{Association of treatment with threshold indicator}: $Z \nip T\mid \Sigma$
\item[R2.] \textit{Independence of guidelines}:$Z \cip \bm{C}\mid (X,\Sigma)$,\\
A weaker version (\ie within strata of $\bm{O}$) is: $Z \cip \bm{U} \mid (X,\bm{O} ,\Sigma)$.
\item[R3.] \textit{Unconfoundedness}: $Y \cip Z\mid (T,\bm{C},X ,\Sigma)$.
\item[R4.] \textit{Continuity}: The expectation of the outcome $E_{\varsigma}(Y\mid X,C,Z,T=t)$ is continuous at $X=x_0$ for $t \in \{0,1\}$ and $\varsigma=\{s,f\}$.
\end{enumerate}

\subsection{MSMM estimator and associated assumptions}\label{sec:msmmassumptions}

We estimate risk ratios in our analysis as these are of primary
interest and most suited to our method of analysis. In the
epidemiological literature odds ratios are usually estimated as they
are easily obtained from logistic regressions. However simple
estimators of causal odds ratios are typically more biased than risk
ratio estimators (\cite{Didelezetal:2010}).

We focus on estimating the Risk Ratio for the Treated (RRT) defined as follows:
\begin{eqnarray}
\frac{E\{E_1(Y\mid Z)\mid T=1\}}{E\{E_0(Y\mid Z)\mid T=1\}} 
\label{eq:RRT}
\end{eqnarray}
which is the binary equivalent of the effect of treatment on the
treated. This quantity can be identified in a fuzzy design provided
additional assumptions, listed below, are satisfied. The RRT can be
interpreted as ratio of the risk for those who were treated relative
to those who were eligible for treatment. 
 In our example we can think about this as follows: statins treatment is made
available to GP-patient pairs (GPPs) around the threshold thus all
these GPPs are eligible for statin prescription. However, of these only
some take up the treatment. The RRT is a comparison of the outcomes
for the patients from GPPs who prescribed statins relative to all the
patients who were eligible and is the binary version of the effect of treatment on the treated(\cite{GenelettiDawid:2010}).

The RRT in (\ref{eq:RRT}) can be identified from observational RD design
data when $Y$, $Z$ and $T$ are binary if in addition to R1-R4 the following assumptions hold (\cite{Didelezetal:2010}).  
\ben
\item[M1:]\textit{Log-linear in t} \\ $\mbox{log}[E_1(Y\mid T=t,Z=z)] - \mbox{log}[E_0(Y\mid T=t,Z=z)]$ is linear in the treatment.
\item[M2:]\textit{No T-Z interaction on the multiplicative
  scale}\\ This assumption is known as the no-effect modification
  assumption (NEM).  \een
  When Assumptions R1-R4 and M1-M2 hold,
  $\Sigma$ is defined as in Section \ref{sec:assumptions} and conditional
  independences A1 
  and A2
  hold, we can use estimators that have been derived in the binary instrumental
  variables literature in the context of the RD design without further
  deriving any results. Thus we obtain the following formula for the
  RRT (\cite{HernanRobins2006,ClarkeWindmeijer:2010}):
\begin{align}
\label{eq:MSMMform}
\mbox{RRT}_{\varsigma} = 1
-\displaystyle\frac{E_{\varsigma}(Y\mid Z=1)-E_{\varsigma}(Y\mid Z=0)}{E_{\varsigma}(Y\bar{T}\mid Z=1)-E_{\varsigma}(Y\bar{T}\mid Z=0)},
\end{align} 
where $\bar{T} = (1-T)$ and $\varsigma = \{s,f\}$. Equivalent
expressions and details of their derivation and interpretation can be
found in \cite{Didelezetal:2010,Abadie:2003,Angrist:2001}. Note that when $\Sigma=s$, (\ref{eq:MSMMform}) reduces to the causal risk ratio. As fuzzy designs are the norm and sharp the exception we drop the subscript $_{\varsigma}$ for notational clarity, so that implicitly $\Sigma=t$.

For the RRT to be above 1, either the numerator or the
denominator must be negative -- but not both. In our context we would
expect to see a positive numerator as we would expect there to be more
individuals with low cholesterol levels above than below the threshold. This is because the threshold and the treatment are positively correlated and
treatment reduces cholesterol. The denominator involves the product of
$Y$ and `no treatment' $\bar{T}=(1-T)$. As we expect statin treatment
to lower cholesterol and be associated with the threshold indicator,
we expect the denominator to be negative. This is reflected in the
results in Section \ref{sec:analysis}, where the RRT is above 1 for both
outcomes and most bandwidths except the smallest for the outcome LDL
below 2.

Assumption M2 requires that whether the LDL cholesterol level is below 2
(resp. 3) does not depend on an interaction term between $T$ (statin
prescription) and $Z$ (whether the risk score exceeds 20\% on the log
scale). If there were an interaction term it would mean that the GPs above and below the
threshold would be different with respect to their ability to predict the outcome. This is unlikely to be the case.

The RRT as show in Equation (\ref{eq:MSMMform}) also identifies the
local causal risk ratio, the binary equivalent of the complier average
causal effect (\cite{FrangakisRubin:2002,Imbens:94}).

\section{Models}\label{sec:models}

The models in Sections \ref{sec:bayes1} and \ref{sec:bayes2} are
embedded in a Bayesian framework. We first obtain a full posterior
MCMC sample for each of the relevant parameters in the models
described in Section \ref{sec:bayes1} and combine these to induce a
posterior sample for the RRT. When we add prior constraints in Section
\ref{sec:bayes2} we sample the RRT directly. From the posterior samples
we easily obtain variances and interval estimates without having to
rely on bootstrap methods or asymptotic arguments, as is the case with
the frequentist estimators.

We present a number of possible models to
estimate the components. We mostly use the same types of models in the
numerator and the same type in the denominator. However, we do mix
different models in the numerator and denominator where we consider
this necessary. Generally speaking we write the estimates for the RRT
as follows: $\mbox{RRT}_{num.denom}$ where \textit{num} indicates the form of
the numerator and \textit{denom} the denominator in the fraction in~(\ref{eq:MSMMform}).

\subsection{Interaction vs Product models}
The denominator of the fraction in the expression for the RRT (henceforth only 
``the denominator'') is given by $E(Y\bar{T}\mid Z = 1) -  E(Y\bar{T}\mid Z = 0)$.
We can break up the individual terms further as follows:
\begin{align}
\label{eq:condprob}
E(Y\bar{T}\mid Z = z) = E(Y \mid \bar{T},Z)E(\bar{T}\mid Z).
\end{align}
In our analysis, we produce estimates for both of these models. We call \textit{interaction} models those which use the formulation on the left hand side of equation (\ref{eq:condprob}) and \textit{product} models those that use the formulation on the right hand side. Our motivation for including analyses with the product of two conditional probabilities as in (\ref{eq:condprob})
is that the data for the product term are sparse (see Supplementary materials). By using $Y$ alone this is mitigated. We also consider zero inflated Poisson regression models (\cite{ZIP}) to address this but results are not substantially different.

\subsection{Poisson regression models}\label{sec:bayes1}

The first set of models we considered estimates all the components in the RRT using Poisson regression models, in line with assumption M1. It is easy to verify that if the same parametric form can be assumed to hold for experimental and observational regimes then a log-linear relationship in $t$ follows for each of the components of (\ref{eq:MSMMform}).

Let $X^*=X-0.2$, we fit Poisson regressions in both the numerator and the denominator:
\begin{align*}
\mbox{Numerator}:=&
\begin{cases}
 y_{i_z} \sim \mbox{Poisson}(\mu_{i_z} )\\
\mbox{log}(\mu_{i_z}) = \alpha_z + \beta_z x_{i_z}^*
\end{cases} 
\mbox{Denominator}:=&
\begin{cases}
y\bar{t}_{i_z}  \sim \mbox{Poisson}(\nu_{i_z})\\
\mbox{log}(\nu_{i_z})  = \delta_z + \gamma_z x_{i_z}^*
\end{cases} 
\end{align*} 
with priors
$\alpha_z,\beta_z,\delta_z,\gamma_z \stackrel{iid}{\sim} \mbox{Normal}(0,100)$, $i_z = 1,\ldots, n_z$ and $ z \in \{0, 1\}$ throughout.

We put relatively vague priors on the regression coefficients. Tighter
priors such as those suggested in \cite{Gelmanetal:2008} have been
considered, but results are not very sensitive to the choice of
prior. As we centre the risk score at the threshold, the parameter of
interest in all the regressions is the exponential of the intercept
term. The posterior MCMC samples of the parameters $\alpha_1$,
$\alpha_0$, $\delta_1$ and $\delta_0$ can be used to characterize $E(Y \mid Z =
1)$, $E(Y \mid Z = 0)$, $E(Y\bar{T}\mid Z = 1)$, and $E(Y\bar{T}\mid Z
= 0)$ respectively and then combined to obtain the posterior sample for the
RRT.
The model described above has the interaction model denominator:
\begin{align*}
\mbox{RRT}_{pois.pois} & = 1 - \frac{\Pi_{pois}}{\Psi_{pois}} \qquad \mbox{where} \\
\Pi_{pois}   = \mbox{exp}(\alpha_1) - \mbox{exp}(\alpha_0)  \qquad & \mbox{and } \qquad \Psi_{pois}  = \mbox{exp}(\delta_1) - \mbox{exp}(\delta_0)
\end{align*} 
We also consider an interaction model where the denominator is based on a flexible Binomial model as used in \cite{Genelettietal:2015}.
In this model the prior information is used to create distance between the two elements in the denominator of the fraction in the
RRT in (\ref{eq:MSMMform}). This often stabilises the results because it pushes the difference in the probability of treatment at the threshold away from zero and thus inflates the fraction in (\ref{eq:MSMMform}). In this case the denominator is defined as
\begin{align*}
 \mbox{Denominator:}\,\,& y\bar{t}_{i_z} \sim \mbox{Binomial}(q_z , n_z )\\
\mbox{priors:}\,\mbox{logit}(q_1) & \sim \mbox{Normal}(-3, 1) \qquad \mbox{and} \qquad  \mbox{logit}(q_0 ) \sim \mbox{Normal}(3, 1) \qquad \mbox{so} \,\, \Psi_{flex} = q_1 - q_0. 
\end{align*} 
This results in the interaction model 
\begin{align*}
\mbox{RRT}_{pois.flex} = 1 - \frac{\Pi_{pois}}{\Psi_{flex}}.
\end{align*}
We now consider the product denominator as follows:
\begin{align*}
\mbox{Denominator.prod}= &
\begin{cases}
y_{i_z}  \sim \mbox{Poisson}(\theta_{iz}) \\
\mbox{log}(\theta_{i_z})  =  \delta_z + \gamma_z x_{i_z} + \kappa_z \bar{t}_{i_z} \\ 
t_{i_z}  \sim \mbox{Binomial}(r_z , n_z)
\end{cases}\\
\mbox{priors:}\,\,& \mbox{logit}(r_1 ) \sim \mbox{Normal}(-3, 1) \qquad \mbox{and } \qquad \mbox{logit}(r_0 ) \sim \mbox{Normal}(3, 1).
\end{align*}
We then combine the conditional probabilities as follows
\begin{align*}
  E(Y\bar{T} \mid Z = z)  = (\delta_z + \kappa_z ) \times r_z,
\end{align*}
as we are interested in the case where both $Y$ and $\bar{T}$ are
equal to 1. Note that we use the Binomial flex model again for the
probability of $\bar{T}$ as this was less variable than regression-based models, in this case. Thus we obtain
$\mbox{RRT}_{pois.prod.flex}$ as follows:
\[
\mbox{RRT}_{pois.prod.flex} = 1 - \frac{\Pi_{pois}}{\Psi_{prod.flex}} 
\]
where 
$\Psi_{prod.flex}  = \mbox{exp}(\delta_1 + \kappa_1 )r_1 - \mbox{exp}(\delta_0 + \kappa_0 )r_0$.

\subsection{Constraints}\label{sec:bayes2}

The RRT can drop below zero if the fraction in
(\ref{eq:MSMMform}) exceeds one (\cite{ClarkeWindmeijer:2010}). We avoid the problem by imposing a priori constraints on
the distribution of the RRT which force the RRT to remain within the acceptable
bounds.

Imposing prior constraints is easy in the Bayesian framework. We put a
Gamma prior on the RRT with most of the mass close to 1, as we do not
want to encourage a large risk ratio. The most straightforward constraint was
to make $\alpha_1$ a function of the other variables above the
threshold so that we could place a prior on the RRT. We could equally
have chosen $\alpha_0$. We write out the changes the model implies to
the priors below:
\begin{align*}
\mbox{RRT}_{pois.pois}  \sim \mbox{Gamma}(3,1) \qquad \mbox{and} \qquad \alpha_1  = \mbox{log}\{(1-\mbox{RRT}_{pois.pois})\Psi_{pois} + \mbox{exp}(\alpha_0)\}
\end{align*}
where $\exp(\alpha_1) - \exp(\alpha_0) =\Pi_{pois}$ and there is of course no prior on $\alpha_1$. Similar
changes can be made to all the RRTs with any of the other models presented in Section \ref{sec:models}. It is also possible to impose constraints on logistic regression based estimates.


\subsection{Generalized Method of Moments analysis}

In order to assess the performance of our Bayesian estimators, we
compared them to some of the most common estimators for binary
outcomes in the IV literature. These include the Generalized Method of
Moments (GMM) MSMM estimator (\cite{Clarkeetal:2015}), the Wald Risk
Ratio (\cite{Didelezetal:2010}) and a final method based on a single
estimating equation (\cite{Burgessetal:2014}). We give a brief
overview and show results in Table \ref{tab:res} of Section
\ref{sec:analysis} for the first of these methods as it outpeformed the other two. Details and results for the other methods can be found in the Supplementary material.

\cite{Clarkeetal:2015} showed that the semi-parametric MSMM estimator in equation (\ref{eq:MSMMform}) can be estimated efficiently by the Generalized Method of Moments (GMM). The moment conditions
\[
\label{eq:mom}
E \left[
\begin{gathered}
Y \exp(-\beta_0X) - \alpha_0 \\
\{Y \exp(-\beta_0X) - \alpha_0\}Z
\end{gathered}
\right] = \left(
\begin{gathered}
0 \\
0
\end{gathered}
\right)
\]
lead to the estimation of $\alpha_0$ and $\beta_0$ by the method of moments.

\section{Simulation study}\label{sec:sim}

We set up a simple simulation study aimed at examining the properties
of the models presented in Section \ref{sec:models} under two levels
of unobserved confounding ($U$), two levels of IV strength ($Z$) and
three causal effects. Thus we performed simulations for 12
Scenarios. We based our simulated data set on the larger data set from
which the one described in Section \ref{sec:rd} and analysed in
Section \ref{sec:analysis} was obtained. Thus we used the original
values for the risk score $X$ and the standardized HDL cholesterol
level as unobserved confounder $U$. The threshold indicator $Z$ was
defined deterministically as $Z_i=0$ if $X_i < 0.2$ and $Z_i=1$ otherwise. For each simulation and in the real data in Section
\ref{sec:analysis} we run the analyses in each of four bandwidths identified by values
$h=\{0.025, 0.05, 0.075, 0.1\}$ and assess the sensitivity of the
results to these changes. 

We generate assigned treatment and outcome in order to get different settings of strength of instrument (Weak or Strong), unobserved confounding (Low or High) and three different risk ratios: 1, 2.11 and 4.48 corresponding to no, low and high causal effect. More details about the simulation mechanism are in the Supplementary material.

We produced exploratory plots like those introduced in Section
\ref{sec:rd} for all the simulated scenarios. Briefly, for high causal
effects all scenarios showed a clear jump at the threshold although it
was smaller in the weak IV high confounding scenario. For the low
causal effects results were variable. No jump was discernible in the
weak IV high confounding scenario but a clear jump was visible in the
strong IV scenarios. Finally for the no-effect scenarios no jump was
visible as expected. A selection of plots can be seen in the Supplementary Materials.


\begin{figure}[!h]
\begin{center}
\includegraphics[scale=.5]{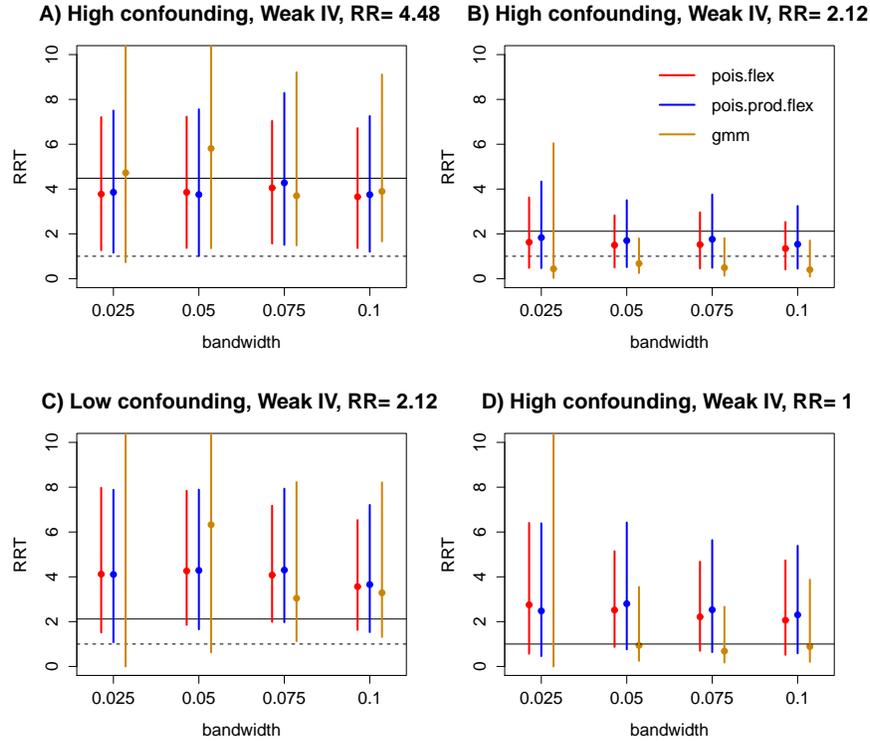}
\end{center}
\caption{\label{fig:SimScen} Comparison of the results in 4 out of 12 simulated scenarios.}
\end{figure}

We obtained results for  Bayesian constrained and unconstrained models
as  well as  a number  of  frequentist estimates  and the  Balke-Pearl
bounds which are available on website \url{xxx} 
In the body of the paper we only present results for the constrained
models as in many scenarios (in particular for small bandwidths, high
confounding and weak instruments) the posterior MCMC sample for the
unconstrained models contained values below zero. In addition, while
convergence was reached for the relevant parameters, there were some
extreme results due to small values in the denominator that
occasionally led to inflated mean estimates. Medians for the
unconstrained models and means for the constrained models were
generally close although the constrained model estimates were
typically smaller. We thought this might be due to the Gamma prior
pulling the RRT in the constrained models towards one. However upon
investigation we saw that results were not sensitive to the choice of
prior.

Figure \ref{fig:SimScen} shows the results for 3 estimators in 4 out of 12 simulated scenarios (full results for all simulated scenarios are available on request). We selected \textit{pois.flex} (in red) and  \textit{pois.prod.flex} (blue). Also we plot the MSMM
estimator obtained using GMM (orange).
For each estimator, a solid coloured line represents the 95\% confidence or
credible interval
 while the dot represents the mean value. The
true risk ratio is shown using a solid horizontal black line, and in case the true effect differs from 1, a dotted line indicates the absence of effect.

In panel A the risk ratio is 4.48. Here the Bayesian model outperforms the others. The mean RRT is closer
to the true value and the intervals are considerably narrower,
especially when small bandwidths are considered. 


In panels B and C the true risk ratio is
equal to $2.11$ and the estimators show mixed behaviour. Results in B are relative to the worst possible scenario,
where confounding is high and the instrument is weak. All
estimators underestimate the real effect and all 95\% intervals
include 1. Bayesian Point estimates are less biased
and their intervals always include the true value. This is not the case for GMM whose intervals do not
include 2.11 for most bandwidths.

Results in C are for the low confounding-weak instrument
scenario. Bayesian estimators show similar behaviour for all values of
$h$, overestimating the true value, but including it within the 95\%
interval. The GMM based method gives wider and misleading intervals
which include the value 1 in the smaller bandwidths $h \in \{0.025,
0.05\}$. 

Finally panel D represents the scenario with high level of
confounding, weak instrument and no effect. With
$h>0.025$ the GMM estimator perform better in terms of both point and
interval estimation however it is very sensitive to the sample
size. In $h=0.025$ the intervals are very wide. The Bayesian estimates
result in very high point estimates but their intervals include 1.

Overall the Bayesian estimators appear to be more robust to IV
strength, confounding levels and size of causal effect. For small bandwidths none of the methods performs particularly well when the risk ratio is small or one, however the Bayesian estimator compares favourably with the competing methods in the borderline cases. A possible reason for the robustness of Bayesian estimators in the extreme scenarios is that continuous information is used in estimating the components of the RDD whereas the GMM (and other frequentist) estimates are based on binary data only. 

\section{Example: Statin prescription}\label{sec:analysis}


\subsection{Data}\label{sec:app}

The data we consider are a subset of The Health Improvement Network
(THIN) data set. THIN are primary care data for over 500 practices in
the UK and include a large number of individual patient, diagnostic
and prescription information. We focus on a subset of 1386 male
patients between 50 and 70 who did not smoke or have diabetes in the
year 2008. We investigate whether statin prescription lowers the LDL
cholesterol to below two and three mmol/L, recommended levels for high
and low risk individuals respectively.


From trials (\cite{HTA}) we know that statins are effective in lowering cholesterol. As LDL
cholesterol tends to decrease quickly within a month of
uptake and our data span the 6 months around the cholesterol
measurement we can use our binary outcomes RD design to determine whether statins result in
people achieving LDL cholesterol targets within a
small time window. Our approach is also useful when we are
interested in whether a drug acts on a relevant marker
of a disease which is easier to measure and is affected quickly by treatment.

\subsection{Preliminary analyses}

Prior to estimating the RRT we investigated whether a Poisson
regression was an appropriate models for the data. Model fit was good
overall and there was no evidence for overdispersion.

In line with recent recommendations regarding what should be presented
in IV analyses (\cite{SwansonHernan:2013}) we performed F-tests to
determine IV weakness for non-linear situations
(\cite{WindmeijerDidelez:2016}) for both binary outcomes (LDL below 2
and below 3). The F-values ranged from 10 (for bandwidth 0.025) to 211
(for bandwidth 0.1) with p-values significant at the 5\% level
throughout. The Balke-Pearl bounds \cite{BalkePearl:1997,PalmerStata:2011} were also in line with our
results.


\subsection{Main analysis}

\begin{table}[h!]
\centering
\small
\begin{tabular}{lllllll}
&\multicolumn{3}{c}{All, LDL$<3$}&\multicolumn{3}{c}{All, LDL$<2$}\\
&Mean&L95&U95&Mean&L95&U95\\
\hline
\multicolumn{7}{l}{$h=0.025$}\\
pois.flex&3.12&1.32&5.62&2.59&0.86&5.57\\
pois.pois&3.98&1.32&8.16&2.96&0.89&6.42\\
pois.prod.flex&3.89&1.11&6.16&2.77&0.77&5.81\\
GMM &3.99&1.30&12.21&8.39&1.49&47.37\\
\hline
\multicolumn{7}{l}{$h=0.05$}\\
pois.flex&3.95&1.51&7.39&2.59&0.99&4.91\\
pois.pois&4.01&1.12&7.93&2.81&0.85&5.92\\
pois.prod.flex&4.17&1.08&8.14&2.64&0.85&5.34\\
GMM&4.53&2.29&8.93&4.49&3.04&29.62\\
\hline
\multicolumn{7}{l}{$h=0.075$}\\
pois.flex&3.76&1.48&7.84&2.63&1.02&4.76\\
pois.pois&4.33&1.47&8.60&3.19&0.95&6.13\\
pois.prod.flex&4.60&1.57&8.60&2.80&1.01&5.33\\
GMM&4.22&2.55&7.03&6.68&3.07&14.56\\
\hline
\multicolumn{7}{l}{$h=0.1$}\\
pois.flex&3.69&1.42&6.82&4.20&1.02&7.13\\
pois.pois&4.02&1.37&7.60&6.82&1.25&4.50\\
pois.prod.flex&3.99&1.36&7.36&2.70&1.31&4.86\\
GMM&3.76&2.60&5.44&7.36&3.73&14.55\\
\hline
\end{tabular}
\caption{The table contains the results for estimates of the constrained RRT as well as the GMM based estimator.\label{tab:res}}
\end{table}

We fitted our models using  \texttt{JAGS} (\cite{JAGS}) with two chains, a burn-in of
10,000 iterations and a further 50,000 iterations. Our posterior
samples were based on the last 1000 iterations. On average each
estimator took 5 minutes to run on a standard PC. Convergence was
reached for all relevant parameters.

Overall the results indicate a positive effect of statin
treatment on LDL cholesterol levels for patients in our sample with a
large (if not universally significant) two to three-fold increase in
the probability of achieving the target LDL cholesterol level within 6
months of prescription for those prescribed relative to those
eligible. This is especially true for the target of reducing the LDL cholesterol level to below 3 mmol/L.

\begin{figure}[!h]
 \begin{center}
\includegraphics[scale=0.4]{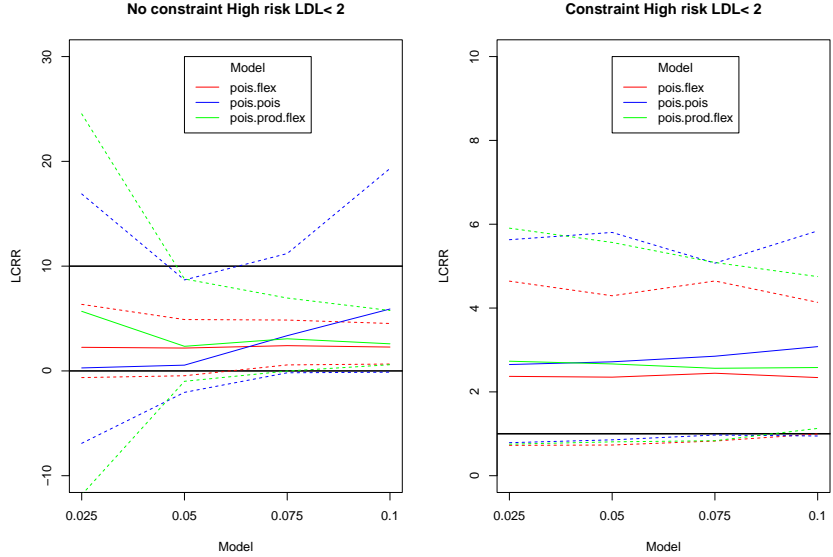}
 \end{center}
 \caption{\label{fig:bandwidth} On the left and right are the means and 95\% credible intervals when the models are unconstrained and constrained respectively. The y-axis have different scales.}
 \end{figure}

Table (\ref{tab:res}) shows for each bandwidth the mean,
upper and lower 95\% interval estimates for the RRT with
constrained models and the GMM.
The plots in Figure \ref{fig:bandwidth} show the RRT estimates for the
unconstrained models in the left and the constrained
models in the right for LDL below 3. We can see that in
particular for the low bandwidths where the data are sparse the
estimates using the unconstrained models are very variable. Similar plots
displaying a similar pattern for the LDL below 2 are given in the
Supplementary Materials. Overall the constrained estimates both reduce
the variability of the estimates and ensure that they remain above zero.

Generally, Bayesian estimates are similar and increase slightly as the bandwidth
increases. The estimates for LDL$<$3 mmol/L are slightly larger than those
for LDL$<2$ and always significant. If you consider that the median
LDL cholesterol for the untreated is approximately 4 then it is easy to see that
to reduce the level to below 2 when statins are thought to reduce
cholesterol by up to 2mmol/L is going to be difficult. Thus it makes
sense that for LDL$<$2 mmol/L the RRT is not significant for small
bandwidths. 


When we compare the results of the GMM methods to ours we see that for
the LDL below 3 scenarios are broadly similar. For the outcome LDL
below 2 however results are different. For all the bandwidths the
means of the GMM are very high and the intervals are wide. Also, from
our simulation studies -- which are based on the same data -- we know
that for small bandwidths (i.e. small sample sizes) the GMM estimators
over-estimate the effects and are less robust in general than the
Bayesian estimates.


\section{Conclusions}\label{sec:conclusion}

In this paper we use the RD design to develop a Bayesian method to
estimate the risk ratio for the treated
(\cite{HernanRobins2006,ClarkeWindmeijer:2010}) which does not stray
below zero. We do this by imposing prior constraints. As this
represents a very strong assumption, we assessed how much results were
affected. Counter to our expectation results from our simulations and
applied example constraining the models to be above zero did not
result in higher RRT estimates, nor were results very sensitive to
prior specification. Specifically varying the values of the Gamma
distribution on the RRT (e.g., by moving the mass further from 1) or
even using a different prior with positive support (i.e., the
log-normal) did not alter results substantially. Instead the
constrained models stabilised the posterior MCMC sample.

The fact that the RTT as identified by (\ref{eq:MSMMform}) has no
built-in safe guard to dropping below zero raises some questions as to
its appropriateness as a risk ratio estimator. It is not easy to
identify causal quantities when all elements are binary and as a
consequence a number of strong assumptions must be met. It is likely
that some will hold only approximately. If we suspect that assumptions
R1-R4 do not hold then there is no point in attempting an RD analysis
or estimating the RRT from these data; however if we think of M1-M2 as
holding only approximately around the threshold then we can view this
as a model misspecification problem and the RRT as an approximation to
the true underlying effect.

Our results compared favourably to those of the generalised method of
moments as well and other frequenstist estimators. In particular they
were more robust to weak instruments, high levels of confounding and
low effects in the simulation studies. They also produced more
credible results in our application.

A further advantage of our approach is that as we model the individual
components of the RRT jointly (possibly subject to constraints) we can
be very flexible in our choice of models. We propose Poisson
regression based models on theoretical grounds and because in the RD
design we can exploit the continuous information in the risk
score. However Poisson regressions do not guarantee that the
coefficients or predictions of the regression behave like
probabilities. While our results do not misbehave in this way it is
something to bear in mind. Due to the flexible approach we have
developled we also have implemented models identical to those we
propose where logs are replaced by logits and exponentials with
expits. Futher we have tried zero-inflated Poisson regression models
and Binomial models. Results are broadly similar especially for the
regression based models. \texttt{JAGS} code for some of these models is given
in the Supplementary materials and results are available on request. It is also feasible to use splines or other semi-parametric models.

There are also a number of papers that focus on estimating causal odds
ratios, notably
\cite{Vansteelandtetal:2011,VansteelandtGoetghebeur:2003,vanderLaanetal:2007}.
While we have not done this here it should be possible to use the
double logistic causal odds ratio estimator in a way similar to how we
use the MSMM risk ratio estimator although additional requirements
(\eg\ specifying a model for $P(Y\mid T,Z)$) need to be taken into
account.

\if1\blind
{
\section*{Acknowledgements}

This research has been funded by a UK MRC grant MR/K014838/1. Approval for this study was obtained from the Scientific Review Committee in August 2014.
} \fi

\if0\blind
{
  
} \fi

\bibliographystyle{plainnat}
\bibliography{binary}
\end{document}